\documentclass[aps]{revtex4-1}
\usepackage{graphicx}
\usepackage{amsmath}
\usepackage{amssymb}
\usepackage{float}
\usepackage{sidecap}
\usepackage{multirow}
\usepackage[caption = false]{subfig}
\usepackage{xcolor}

\usepackage{braket}
\usepackage{tabularx,booktabs}
\newcolumntype{Y}{>{\centering\arraybackslash}X}
\usepackage[margin=1in]{geometry} 

\begin{document}
\title{\bf Spectroscopy of excited charmed mesons}
\author{Mohammad H. Alhakami}
\affiliation{Department of Physics and Astronomy, College of Science, King Saud University,
P. O. Box 2455, Riyadh 11451, Saudi Arabia}
\affiliation{Nuclear Science Research Institute, KACST, P.O. Box 6086, Riyadh 11442, Saudi Arabia}
\affiliation{School of Physics and Astronomy,
University of Glasgow, Glasgow, G12 8QQ, United Kingdom}
\date{\today}
\begin{abstract}
We derive mass formulas
for the $P$-wave orbitally excited $D^*_{0(s)}$, $D^\prime_{1(s)}$, $D_{1(s)}$, and $D^*_{2(s)}$ heavy charmed mesons including  
all effects from one-loop corrections that contribute 
at leading order in chiral expansion.
In our formalism, the effects to first order in $m_q$, where $m_q$ is the light quark mass, and to first order in $m^{-1}_c$, where $m_c$ is the charm quark mass,
and $m_q/m_c$ terms
are considered. The experimental and  
lattice QCD results on the charmed meson spectra are employed
to fix the large number of counterterms
appearing in the effective chiral Lagrangian used in this work. This allows us to test the validity of perturbative expansion of our theory.  The results
presented in the current paper are useful 
to other applications
of excited charmed and bottom meson systems. 
\end{abstract}
\pacs{Valid PACS appear here}
\maketitle
\onecolumngrid

\section {Introduction}
The spectra and decays of the
charmed mesons have been studied extensively, for a review, see \cite{rev}. 
The properties of these mesonic bound states, which contain heavy quarks, are almost described 
using heavy quark symmetries. 
In the heavy quark limit, heavy quark spin decouples
from the dynamics of QCD and the spin and parity, $j^P_l$, of the light degrees of freedom (light antiquarks and gluons) are used to 
classify degenerate charmed meson states into spin doublets. 
The low-lying states form the following
heavy spin doublets,
\begin{equation}\label{doublets}
\underbrace{\left[D_q(0^-),D^*_q(1^-)\right]}_{j^P_l=\frac{1}{2}^- (L=0)}, 
\underbrace{\left[D^*_{0q}(0^+),D^\prime_{1q}(1^+)\right]}_{j^P_l=\frac{1}{2}^+ (L=1)},
\underbrace{\left[D_{1q}(1^+),D^*_{2q}(2^+)\right]}_{j^P_l=\frac{3}{2}^+ (L=1)},
\end{equation} 
where $q$ is the SU(3) index.
The strong interactions of these heavy mesons with soft pseudo-Goldstone bosons ($\pi$, $K$, and $\eta$) are constrained by 
chiral symmetry. The formal approach to employ these two approximate symmetries 
of QCD when investigating
the properties of mesons containing heavy quarks is an effective field theory. 
The effective field theory that describes the low-energy strong interactions of heavy mesons and light pseudo-Goldstone bosons is heavy meson chiral perturbation theory (HMChPT).

Within this framework, an effective Lagrangian, which obeys chiral and heavy quark symmetry constraints, is built to analyze the spectroscopy of the ground-state $j^P_l=\frac{1}{2}^-$ doublets at the heavy quark limit (Refs. \cite{n1,n2,n3,n4,n5,n6}), including corrections to first order in  light quark and charm quark masses (Refs. \cite{cdgn12,n7,4}). The authors of Ref. \cite{ms05} have extended the applications of this effective theory to study the masses 
of the ground-state ($j^P_l=\frac{1}{2}^-$ doublets) and the lowest excited-state ($j^P_l=\frac{1}{2}^+$ doublets) charmed mesons. In their studies, the corrections from  leading order chiral and heavy quark symmetry violating terms and one-loop effects from couplings within and between charmed mesons
that form  $j^P_l=\frac{1}{2}^-$ and $j^P_l=\frac{1}{2}^+$ heavy spin doublets
have been considered. However, the loop effects from the coupling of these states to the higher excited-state
$j^P_l=\frac{3}{2}^+$ doublets are not calculated in \cite{ms05}. As emphasized by the authors of \cite{ms05}, the virtual loop effects from these higher excited states ($j^P_l=\frac{3}{2}^+$ doublets) are crucial to the physics of $j^P_l=\frac{1}{2}^+$ doublets. This is because $j^P_l=\frac{1}{2}^+$ and $j^P_l=\frac{3}{2}^+$ doublets are separated by nearly $\leqslant$ 130 MeV, and  their coupling is at leading order in derivative chiral expansion. The single  pion transition between excited-state $j^P_l=\frac{3}{2}^+$ doublets and the ground-state $j^P_l=\frac{1}{2}^-$ doublets  proceeds through $d$ waves and is hence
suppressed by one derivative in the effective Lagrangian \cite{FalkLuke}.

In the present paper, the virtual loop effects from the higher excited-state $j^P_l=\frac{3}{2}^+$ doublets to the masses of the lowest excited-state $j^P_l=\frac{1}{2}^+$ doublets 
are calculated. We also use the third-order chiral Lagrangian, which includes the relevant excited charmed mesons as explicit degrees of freedom,
to derive mass expressions for the excited-state $j^P_l=\frac{3}{2}^+$ doublets
including all leading loop effects and 
corrections due to chiral and heavy quark symmetry breaking.
The mass formulas for the excited-state
$j^P_l=\frac{1}{2}^+$ and $j^P_l=\frac{3}{2}^+$ doublets contain a large number of unknown 
parameters that cannot be determined uniquely from 
experimental measurements on the meson spectrum alone.
We, therefore, follow the approach employed in  our previous work \cite{Alhakami,LECs} to fit these unknown 
counterterms. It is based on (i) 
reducing the number of unknown parameters by grouping them into a number of linear combinations that is equivalent to the number of observed charmed states and (ii) 
using experimental information on masses and couplings to evaluate the loop functions, which makes the fit linear. The privilege of feeding loop integrals with physical masses is to put the threshold of decaying particles in the correct place.
The chirally symmetric terms appearing in some linear combinations 
can be disentangled
from SU(3) symmetric terms  
using lattice QCD results on the charmed meson spectrum. 
The lattice QCD work undertaken by the authors of \cite{lattice} provides sufficient information for performing this task.
By fitting counterterms, our mass expressions can be applied to investigate various mass splittings within excited charmed mesons and their analog bottom mesons.
 
This paper is organized as follows. The effective Lagrangian formalism we use is presented in Sec. II. In Sec. III, the mass formulas for excited charmed meson $j^P_l=\frac{1}{2}^+$ and $j^P_l=\frac{3}{2}^+$ doublets are given. 
They contain a large number of unknown counterterms, which can be fixed using empirical and lattice information on masses and coupling constants. Sec. IV explains the fitting method and draws conclusions on the results and validity of ChPT in the heavy light sector.
\section{effective chiral Lagrangian}
Before writing down the relevant effective chiral
Lagrangian, let us first introduce the fields representing the light pseudo-Goldstone
and heavy charmed mesons and show how they change under chiral $SU(3)_L\times SU(3)_R$ and heavy quark $SU(2)_s$ symmetry transformations; we refer the interested reader to Refs. \cite{Falk,cas97,MehenFalk,FalkLuke,ms05,PCh,4,tensor,sch03}. 

The pseudo-Goldstone octet is incorporated into the $3\times3$ unitary matrix
$U(x)=\exp({\rm i}2\phi(x)/f)$, where  $\phi(x)$ is given by %\cite{PCh}
\begin{equation}
\phi(x)=\sum_{i=1}^8 \frac{\lambda_i \phi_i(x)}{2}=\frac{1}{2}\left (\begin{array}{ccc}\pi^0+\frac{1}{\sqrt{3}}\eta&\sqrt{2}\pi^+&\sqrt{2}K^+ \\ \sqrt{2}\pi^- & -\pi^0+\frac{1}{\sqrt{3}}\eta&\sqrt{2}K^0\\
\sqrt{2}K^-&\sqrt{2}\bar{K}^0&-\frac{2}{\sqrt{3}}\eta \end{array}\right),
\end{equation}
where $\lambda_i$ are the Gell-Mann matrices and $f$ is the pion decay constant, $f=92.4$ MeV.
The field $U$ transforms linearly under chiral symmetry, $U \rightarrow R U L^\dag$, where
$R$ and $L$ represent global elements of 
$SU(3)_R~\text{and}~ SU(3)_L$, respectively. To describe the interactions of pseudo-Goldstone bosons with matter fields representing, in our case, the heavy charmed mesons, it is convenient to introduce the coset field $u(x)=\sqrt{U(x)}$. The field $u$ transforms nonlinearly under chiral symmetry, $u \rightarrow R u K^{-1}$, where the $SU(3)$-valued function $K$ is given by $K(L,R,U)=(\sqrt{RUL^\dag})^{-1}R\sqrt{U}$. 

The pseudo-Goldstone bosons derivatively couple to heavy mesons through the vector and axial vector combinations,
\begin{equation}
\begin{split}
\mathcal{V}^\mu &=\frac{1}{2}(u^+\partial^\mu u+u\partial^\mu u^+)=\frac{1}{2f^2}[\phi,\partial^\mu\phi]+O(\phi^4),\\
\mathcal{A}^\mu&=\frac{i}{2}(u^+\partial^\mu u-u\partial^\mu u^+)=-\frac{1}{f}\partial^\mu\phi+O(\phi^3).
\end{split}
\end{equation}
Under the unbroken $SU(3)_{L+R}$ flavor symmetry, the $\mathcal{A}^\mu$ and $\mathcal{V}^\mu$ fields transform homogeneously, $\mathcal{A}^\mu \rightarrow K \mathcal{A}^\mu K^\dag$, and inhomogeneously, $\mathcal{V}^\mu\rightarrow K \mathcal{V}^\mu K^\dag+K\partial^\mu K^\dag$.

The heavy meson fields representing the components of heavy spin doublets shown in Eq.~\eqref{doublets} are incorporated into the following $4\times4$ matrices:
\begin{equation}
\begin{split}
H_q&=  \frac{1+v\llap/}{2\sqrt{2}}\left(D^{*\mu}_q\gamma_\mu-D_q\gamma^5\right),\\
S_q&= \frac{1+v\llap/}{2\sqrt{2}}\left(D^{\prime\mu}_{1q}\gamma_{\mu}\gamma^5-D^*_{0q}\right),\\
T^\alpha_q&= \frac{1+v\llap/}{2\sqrt{2}}\left(D^{*\alpha \mu}_{2q}\gamma_{\mu}-D_{1q\mu}\sqrt{\frac{3}{2}}\gamma^5 [g^{\alpha \mu}-\frac{1}{3}\gamma^\mu(\gamma^\alpha-v^\alpha)] \right),
\end{split}
\end{equation}
where the various operators annihilate heavy mesons of four-velocity $v$ with quark content $Q\bar{q}$ and the subscript $q$ stands for light quark flavor. Here, we use the notation employed
in \cite{MehenFalk} to define the fields for the charmed meson states.
In our approach, we have chosen to define the 
nonrelativistic meson fields $D_q$, $D^{*\mu}_q$, $D^*_{0q}$, $D^{\prime\mu}_{1q}$, $D^{\mu}_{1q}$, and $D^{*\alpha\mu}_{2q}$ in four dimensions to maintain the heavy quark symmetry
at the quantum level \cite{Alhakami}. In the current work, we neglect the possible mixing between the axial-vector
$D_1^{\prime\mu}$ and $D_1^\mu$ charmed meson states. 

The above fields are normalized as follows:
\begin{equation}
\begin{split}
\bra{0} D^{(*)}_{(0)q} \ket{Q\bar{q}(0^{-(+)})}&=1,~\bra{0} D^{*\mu}_q \ket{Q\bar{q}(1^-)}=\epsilon^\mu, ~
\bra{0} D^{(\prime) \mu}_{1q} \ket{Q\bar{q}(1^+)}=\epsilon^{(\prime)\mu}, ~
\bra{0} D^{*\mu\nu}_{2q} \ket{Q\bar{q}(2^+)}=\epsilon^{\mu\nu},\\
\end{split}
\end{equation}
where $\epsilon^{\mu}$ ($\epsilon^{\mu\nu}$) is the polarization vector (tensor) of the initial state. 
The vector and tensor polarizations are normalized as
$\epsilon\cdot\epsilon=-1$ and $\epsilon^{\mu\nu}\epsilon_{\mu\nu}=1$, respectively, and satisfy
$v^\mu \epsilon_\mu=0$, $\epsilon_{\mu\nu}=\epsilon_{\nu\mu}$, $v^\mu \epsilon_{\mu\nu}=v^\nu \epsilon_{\mu\nu}=0$, and $\epsilon_{\mu\nu}g^{\mu\nu}=\epsilon^\mu_\mu=0$. 

The velocity-dependent superfields $H_q$, $S_q$, and $T^\alpha_q$ transform as doublets under heavy quark symmetry $SU(2)_s$ and as antitriplets under the unbroken flavor $SU(3)_{L+S}$. Their complex conjugates are defined as $\bar{H}_q=\gamma^0 H^+_q \gamma^0,~\bar{S}_q=\gamma^0 S^+_q \gamma^0$, and $\bar{T}^\alpha_q=\gamma^0T^{+\alpha}_q\gamma^0$.

Having introduced the field operators for the
light pseudo-Goldstone
and heavy charmed meson particles, we are now 
in a position to 
present the most relevant pieces of the chiral Lagrangian.
We begin by writing the lowest-order Lagrangian for the pseudo-Goldstone bosons, 
\begin{equation}\label{1}
\mathcal{L}_m=\frac{f^2}{4}<\partial_\mu U\partial^\mu U^\dagger>+\frac{f^2\,B_0}{2}<m_q\, U^{\dag }+U\,m_q^{\dag}>,
\end{equation}
where $<...>$ means the trace and the factor $B_0$ is related to  the quark condensate 
of light quark flavors and the pion decay constant. The quantity $m_q$ is the light quark mass matrix, $m_q=\mathrm{diag}(m_u,m_d,m_s)$. We work in the isospin limit, $m_u=m_d=m_n$ and $m_q=\mathrm{diag}(m_n,m_n,m_s)$, where the subscripts $n$ and $s$ denote nonstrange
and strange light quark flavors, respectively.

The kinetic piece of the effective Lagrangian
describing heavy fields is
\begin{equation}
\begin{split}
{\mathcal{L}}_{\mathrm{kin}}=&-<\bar{H_a}\left(i v\cdot D_{ba} 
-\delta_H \delta_{ab}\right) H_b>
+<\bar{S}_a\left(i v\cdot D_{ba} 
-\delta_S \delta_{ab}\right) S_b>
+<\bar{T}^\alpha_a\left(i v\cdot D_{ba} 
-\delta_T \delta_{ab}\right) T_{\alpha b}>,
\end{split}
\end{equation}
where $\delta_A$, $A \in \{H,S,T\}$, represents the residual mass of the sector $A$ and the covariant derivative is given by
$D^\mu_{ba}=\delta_{ba}\partial^{\mu}+\mathcal{V}^\mu_{ba}$.
The free propagators for the heavy fields are 
\begin{equation}
\begin{split}
\text{scalar meson:}&~~\frac{i}{v.k},\\
\text{vector meson:}&~~\frac{-i(g^{\mu \nu}-v^{\mu} v^{\nu})}{v.k},\\
\text{tensor meson:}&~~\frac{i}{v.k}\frac{1}{2}\left(
(g^{\mu\nu}-v^\mu v^\nu)(g^{\rho \sigma}-v^\rho v^\sigma)+(g^{\mu\sigma}-v^\mu v^\sigma)(g^{\nu \rho}-
v^\nu v^\rho)-\frac{2}{3}(g^{\mu\rho}-v^\mu v^\rho)(g^{\nu \sigma}-v^\nu v^\sigma)\right).
\end{split}
\end{equation}

We are interested in the low-energy transitions between heavy mesons with a single pseudo-Goldstone bosons. The interactions between states in the same heavy spin doublets are governed by the leading order Lagrangian 
\begin{equation}
 \begin{split}\label{L0}
{\mathcal{L}}^{1}_{\mathrm{int}}=&g <\bar{H}_a H_b {\mathcal{A}\llap/}_{ba} \gamma^5>
+g^{\prime}<\bar{S}_a S_b {\mathcal{A}\llap/}_{ba}\gamma^5>
+g^{\prime \prime} <\bar{T}^\alpha_a T_{\alpha b} {\mathcal{A}\llap/}_{ba} \gamma^5>,
 \end{split}
\end{equation}
where the dimensionless quantities $g$, $g^\prime$, and 
$g^{\prime\prime}$ represent the coupling constants that measure the strengths of strong transitions between charmed states that form  $\frac{1}{2}^-$, $\frac{1}{2}^+$, and $\frac{3}{2}^+$ heavy quark spin doublets, respectively. These coupling constants can be measured experimentally.
The lowest-order interaction Lagrangian that describes the strong transitions between doublets
with a soft single pseudo-Goldstone bosons is given by 
\begin{equation}
 \begin{split}\label{L1}
{\mathcal{L}}^{2}_{\mathrm{int}}=&h <\bar{H}_a S_b {\mathcal{A} \llap/}_{ba} \gamma^5>+
h^\prime<\bar{S}_a T^\mu_b {\mathcal{A}}_{\mu ba} \gamma^5 >+\text{H.c.}\\
 \end{split}
\end{equation}
The strong transitions between $\frac{3}{2}^+$ and $\frac{1}{2}^-$ spin doublets proceed through $d$ waves and are hence  suppressed by one derivative in the chiral Lagrangian. 
For the interactions between doublets, we only consider
the leading contributions given in Eq.~\eqref{L1}.

The other terms in the effective chiral Lagrangian needed are the following higher order mass counterterms
\begin{equation}
\begin{split}
{\mathcal{L}}^{\text{mass}}=&-\frac{\Delta_H}{8}<\bar{H}_a\sigma^{\mu \nu}H_a\sigma_{\mu\nu}>+a_H <\bar{H}_a H_b> m^{u}_{ba} + \sigma_H <\bar{H}_a H_a> m^{u}_{bb}\\
&-\frac{\Delta^{(a)}_H}{8}<\bar{H}_a\sigma^{\mu\nu} H_b \sigma_{\mu\nu}>m^u_{ba}
-\frac{\Delta^{(\sigma)}_H}{8}<\bar{H}_a \sigma^{\mu\nu} H_a\sigma_{\mu\nu}>m^u_{bb}\\
&+\frac{\Delta_S}{8}<\bar{S}_a\sigma^{\mu \nu}S_a\sigma_{\mu\nu}>-a_S <\bar{S}_a S_b> m^u_{ba} - \sigma_S <\bar{S}_a S_a> m^u_{bb}\\
&+\frac{\Delta^{(a)}_S}{8}<\bar{S}_a\sigma^{\mu\nu} S_b \sigma_{\mu\nu}>m^u_{ba}
+\frac{\Delta^{(\sigma)}_S}{8}<\bar{S}_a \sigma^{\mu\nu} S_a\sigma_{\mu\nu}>m^u_{bb}\\
&+ \frac{3}{16}\Delta_T <\bar{T}^\alpha_a\sigma^{\mu \nu}T_{\alpha a}\sigma_{\mu\nu}>- a_T <\bar{T}^\alpha_a T_{\alpha b}> m^u_{ba}-\sigma_T <\bar{T}^\alpha_a T_{\alpha a}> m^u_{bb}\\
&+\frac{3}{16} \Delta^{(a)}_T <\bar{T}^\alpha_a\sigma^{\mu\nu} T_{\alpha b} \sigma_{\mu\nu}> m^u_{ba}+\frac{3}{16}\Delta^{(\sigma)}_T <\bar{T}^\alpha_a \sigma^{\mu\nu} T_{\alpha a}\sigma_{\mu\nu}> m^u_{bb}. \label{Mass}
\end{split}
\end{equation}
where $\Delta$ is the hyperfine operator and $m^u_{ab}$ is the mass matrix, which breaks chiral symmetry, and it is defined as $m^{u}_{ba}=\frac{1}{2}\left( u \, m_q u+ u^\dag m_q u^\dag \right)_{ba}$.
The factors $a$, $\sigma$, $\Delta^{(a)}$, and $\Delta^{(\sigma)}$ are dimensionless coefficients.

\section{$P$-wave charmed meson masses}
\begin{figure}[h!]
\begin{center}
\includegraphics[width = 5in]{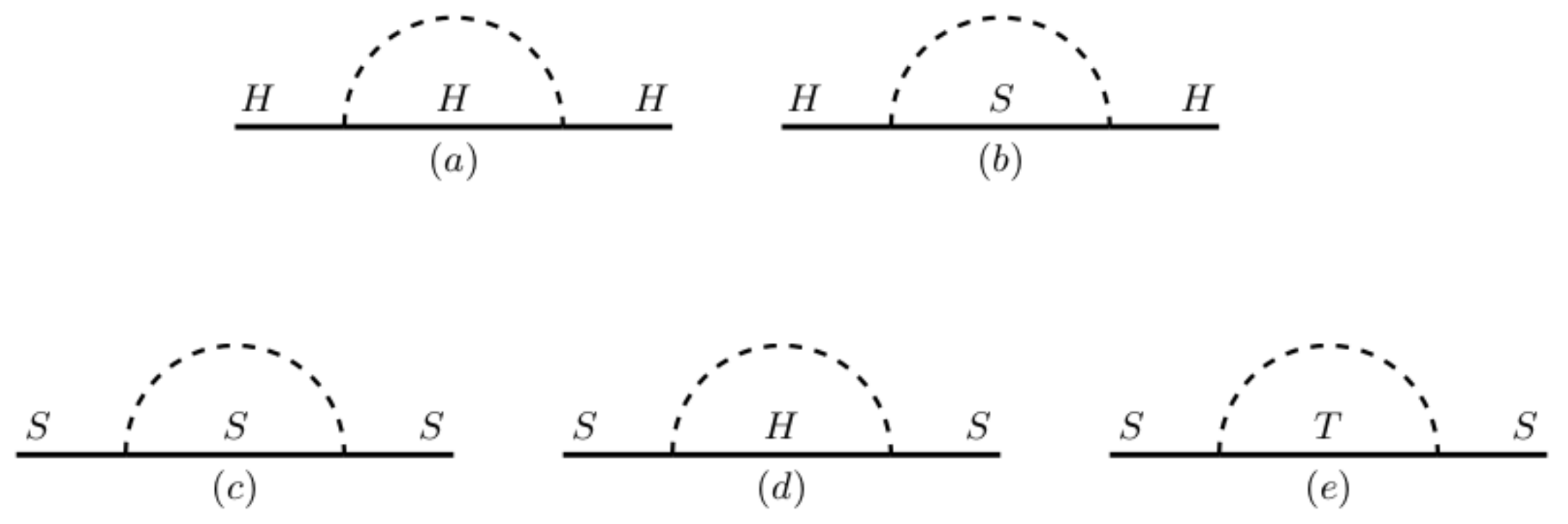}
\caption{Feynman diagrams shown in (a) and (b) represent the self-energy of the $H$ field and those shown in (c)--(e) represent the self-energy of the $S$ field. The dashed line represents the pseudo-Goldstone bosons: $\pi$, $K$, and $\eta$. The loop effects shown in
(a)--(d) have been calculated in \cite{ms05}. The virtual loop effects from the diagram shown in (e) are calculated in the present paper.}\label{fig1}
\end{center}
\end{figure}

The authors of Ref. \cite{ms05} have
used HMChPT to derive the mass formulas for the charmed meson states that form members of the $j^P_l=\frac{1}{2}^-$ and $\frac{1}{2}^+$ spin doublets. 
They expressed the masses up to third order in the chiral expansion including
one-loop corrections and leading heavy quark and chiral symmetry violating terms.
The one-loop graphs they calculated are shown in Figs.~\ref{fig1} (a)-\ref{fig1}(d).
However, the contributions from the leading one-loop graph in
Fig.~\ref{fig1} (e) have not been considered in \cite{ms05}. 
As stated above, these loop effects are important to the physics
of the $j^P_l=\frac{1}{2}^+$ spin doublets; i.e., they contribute at leading order to the interaction Lagrangian as
shown in Eq.~\eqref{L1} [see Eqs.~\eqref{L0} and \eqref{L1}]. 
\begin{figure}[h!]
\begin{center}
\includegraphics[width = 3.5in]{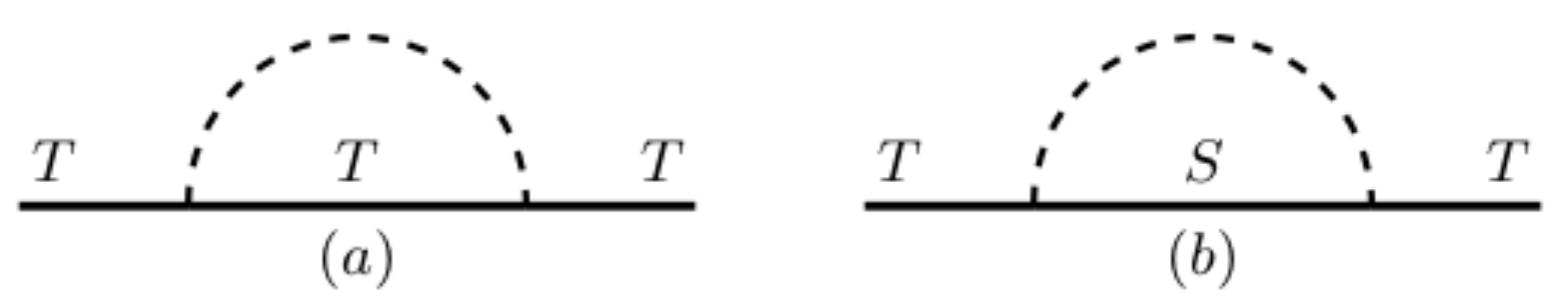}
\caption{The one-loop self-energy diagrams for the $T$ field. The notation is the same as in Fig.~\ref{fig1}.}
\label{fig2}
\end{center}
\end{figure}

The current paper is devoted to extending the applications of HMChPT to 
the spectroscopy of the excited $P$-wave charmed mesons
that form members of the $j^P_l=\frac{1}{2}^+$ and $\frac{3}{2}^+$ spin doublets. 
The missing one-loop corrections represented in Fig.~\ref{fig1} (e) 
are calculated and the mass expressions for the excited states belonging to $\frac{3}{2}^+$ spin 
doublets  are derived up to third order in the chiral expansion including leading 
one-loop corrections (Fig.~\ref{fig2}) and corrections due to breaking of chiral and heavy symmetry. 

Using the effective Lagrangian presented in the previous section, we write down the mass expressions
for all the $P$-wave orbitally excited charmed mesons:
\begin{equation}
\begin{split}\label{masses}
m^r_{D^*_{0q}}&= \delta_S+a_S m_q+\sigma_S \overline{m}-\frac{3}{4}(\Delta_S+\Delta^{(a)}_S m_q+\Delta^{(\sigma)}_S \overline{m})+\Sigma_{D^*_{0q}},\\[1ex]
m^r_{D^\prime_{1q}}&= \delta_S+a_S m_q+\sigma_S \overline{m}+\frac{1}{4}(\Delta_S+\Delta^{(a)}_S m_q+\Delta^{(\sigma)}_S \overline{m})+\Sigma_{D^\prime_{1q}},\\[1ex]
m^r_{D_{1q}}&= \delta_T+a_T m_q+\sigma_T \overline{m}-\frac{5}{8}(\Delta_T+\Delta^{(a)}_T m_q+\Delta^{(\sigma)}_T \overline{m})+\Sigma_{D_{1q}},\\[1ex]
m^r_{D^*_{2q}}&= \delta_T+a_T m_q+\sigma_T \overline{m}+\frac{3}{8}(\Delta_T+\Delta^{(a)}_T m_q+\Delta^{(\sigma)}_T \overline{m})+\Sigma_{D^*_{2q}},\\
\end{split}
\end{equation}
where $m_D^r$ defines the residual masses that are measured with respect to some reference mass of $O(m_Q)$
and $\overline{m}=2m_n+m_s$. 
The symbol $\Sigma_D$ represents the one-loop corrections, which appear at leading order in the chiral expansion, to the excited $D$ meson masses. The one-loop effects are shown in Figs.~\ref{fig1} (c)-\ref{fig1}(e) and Fig.~\ref{fig2}, 
and their explicit expressions are given in the Appendix. 

The theory is a double expansion in 
 $\Lambda_{\mathrm{QCD}}/m_Q$ and $Q/\Lambda_\chi$, where 
$m_Q$ and $Q$ represent the heavy quark mass (charm quark in the case of the charmed mesons) and 
low-energy scales in the theory ($Q\sim m_\pi,m_K,m_\eta$), respectively \cite{ms05}.
The quantity $\Lambda_\chi$ defines the chiral symmetry breaking scale, $\Lambda_\chi=4\pi f \approx 1.5$ GeV.
Based on the power counting rules, the coefficients 
$\delta, \Delta, \Delta^{(a)}, \Delta^{(\sigma)}$ 
scale as $Q$. The terms with light quark mass $m_q \propto m^2_\pi  \sim Q^2$, and hence,
$\overline{m}  \sim Q^2$. The loop functions $\Sigma_D$ scale as $Q^3$.

One can link the terms appearing in the above mass formula to the observed charmed meson spectrum. In the mass expansion [Eq.~\eqref{masses}], terms with $\delta_A$ (chirally symmetric, at order $Q$) and $\sigma_A$ [$SU(3)$ symmetric, at order $Q^2$] coefficients  give the same contributions to charmed meson masses in sector $A$. The $SU(3)$ mass splittings within charmed mesons are due to $a_A$ (at order $Q^2$) 
and $\Delta_A^{(a)}$ (at order $Q^3$)
in our mass expansion.
Terms containing $\Delta_A$, $\Delta_A^{(\sigma)}$, and $\Delta_A^{(a)}$ give rise to chirally symmetric (at order $Q$), chiral symmetry breaking (at order $Q^3$), and $SU(3)$ symmetric breaking (at order $Q^3$) hyperfine splittings, respectively. 
Therefore, by fitting these unknown coefficients, one can use the theory to calculate several mass splittings in the heavy-light meson systems.
\section{Results and conclusion}
To make the theory more predictive, the unknown counterterms appearing in Eq.~\eqref{masses} must be determined using experimental information on charmed meson masses and coupling constants.
However, as their numbers exceed the number of observed spectra, a unique fit for them from utilizing a nonlinear fit is impossible \cite{ms05}. 
Here, we follow the method employed in \cite{Alhakami,LECs} to determine their unique values 
using experimental and lattice information on masses and couplings. 
To this end, let us begin by introducing \cite{Alhakami}
\begin{align}\label{cof}
&\eta_A=\delta_A+(\frac{a_A}{3}+\sigma_A)\, \overline{m},~~~~\xi_A=\Delta_A+(\frac{\Delta^{(a)}_A}{3}+\Delta^{(\sigma)}_A)\, \overline{m},\\[2ex]\label{mm2}
&L_A=(m_s-m_n)\,a_A,~~~~~~~~~F_A=(m_s-m_n)\,\Delta^{(a)}_A,
\end{align}
where $A\in\{S,T\}$. The combinations $\eta_A$ and $\xi_A$ ($L_A$ and $F_A$) preserve (violate) $SU(3)$ flavor symmetry.
The combinations $\xi_A$ and $F_A$ contain the operators
$\Delta_A$, $\Delta^{(a)}_A$, and $\Delta^{(\sigma)}_A$, which break $SU(2)_s$
heavy quark spin symmetry.
The quantity $m_n$ ($m_s$) represents the mass of the nonstrange (strange) light quark. Our one-loop formulas given in Eq.~\eqref{masses} can be expressed in terms of the above defined parameters as
\begin{equation}
\begin{split}\label{rmasses}
&m^r_{D^*_{0q}}=\eta_S-\frac{3}{4}\xi_S+\alpha_q L_{S}-\beta_q F_{S}+\Sigma_{D^*_{0q}},\\[1ex]
&m^r_{D^\prime_{1q}}=\eta_S+\frac{1}{4}\xi_S+\alpha_q L_{S}+\frac{1}{3} \beta_q F_{S}+\Sigma_{D^\prime_{1q}},\\[1ex]
&m^r_{D_{1q}}=\eta_T-\frac{5}{8}\xi_T+\alpha_q L_{T}-\frac{5}{6} \beta_q F_{T}+\Sigma_{D_{1q}},\\[1ex]\
&m^r_{D^*_{2q}}=\eta_T+\frac{3}{8}\xi_T+\alpha_q L_{T}+\frac{1}{2}\beta_q  F_{T}+\Sigma_{D^*_{2q}},
\end{split}
\end{equation}
where the numerical values of the light flavor coefficients $\alpha_q$ and $\beta_q$ are as follows: $\alpha_n=-1/3$, $\alpha_s=2/3$, $\beta_n=-1/4$, $\beta_s=1/2$.
 Now, the number of unknown parameters
 in Eq.~\eqref{rmasses}
 equals the number of charmed meson states. By using physical values of masses and couplings in evaluating loop integrals, the one-loop pieces become constant, and hence one can extract the numerical values of  parameters $\eta_A$, $\xi_A$, $L_A$, and $F_A$
 when fitting the above mass expansion to the observed spectrum. From fit results, one can only fix $SU(3)$-violating coefficients $a_A$ and $\Delta^{(a)}_A$; see Eq.~\eqref{mm2}. However, the other coefficients ($\delta_A$, $\sigma_A$, $\Delta_A$, $\Delta^{(\sigma)}_A$) cannot be fixed using an experimental fit alone; see Eq.~\eqref{cof}. 
To extract them, lattice results on charmed meson masses evaluated at different quark masses (nonphysical pion masses) are needed. 
By fitting the lattice data to the above mass expressions, one can extract parameters $\eta_A$, $\xi_A$, $L_A$, and $F_A$ at different light quark masses. Having determined  $\eta_A$ and $\xi_A$ at physical
(by fitting to experimental data) and nonphysical (by fitting to lattice
data) light quark masses, one can then use a constrained fitting procedure to
fix the coefficients  $\delta_A$, $\sigma_A$, $\Delta_A$, and $\Delta^{(\sigma)}_A$.

Experimental measurements on the charmed meson masses used in this work \cite{pdg12,19} are given in Table \ref{tab:1}. The ground states enter the loop functions that contribute to the masses of the excited-state $\frac{1}{2}^+$ spin doublets; see the Appendix.
In our fit, we use experimentally determined masses: six from the nonstrange sector and six  from the strange sector.
For the nonstrange sector, we take the isospin limit of the 
well-determined masses and use the masses of the excited charmed mesons $D^{0\prime}_1$ and $D_0^{*+}$, which are reported with relatively small uncertainties. We use the following physical values: $m_n=4$ MeV, $m_s=130$ MeV, 
$m_{\pi}=140$ MeV, $m_K = 495$ MeV, and $m_\eta = 547$ MeV. 
For the couplings, the values $g=0.64\pm0.075$ and $h=0.56\pm0.04$, which have been measured from strong decays of the charmed mesons \cite{cdgn12}, are used. The coupling constants $g^\prime$, $g^{\prime\prime}$, and $h^\prime$ are experimentally unknown. However, we use the computed lattice QCD value for $g^\prime=-0.122(8)(6)$ \cite{glattice} and $0.5$ for  $g^{\prime\prime}$ and $h^\prime$. For the normalization scale, we use $\mu=1$ GeV.
 
 \begin{table}[h!]
\def\arraystretch{2}
\begin{center}
\caption{The listed charm meson states are used in our fitting; for details please refer to the text. 
The spin and parity of the light degrees of freedom 
$j^P_l$ are used to classify these heavy mesons; see 
Eq.~\eqref{doublets}. The angular momentum and parity of the meson are represented  by $J^p$. We take all masses from the Particle Data Group \cite{pdg12}
except the mass of $D_1^{0\prime}$, which is reported by the Belle collaboration \cite{19}.}
\begin{tabular}{cccccccc}
\hline\hline
$j^P_l$        &$J^P$&c$\bar{u}$     &$M$(MeV)   &c$\bar{d}$       &$M$(MeV)    & c$\bar{s}$       &$M$(MeV)   \\ \hline
               
$\frac{3}{2}^+$&$2^+$&$D_2^{*0}$     &$2460.7(4)$ &$D^{*+}_{2}$     &$2465.4(1.3)$&$D^{*+}_{s2}$     &$2569.1(8)$\\
$\frac{3}{2}^+$&$1^+$&$D_1^{0}$      &$2420.8(5)$ &$D^+_1$&$2423.2(2.4)$        &$D^{+}_{s1}$      &$2535.11(6)$\\  
$\frac{1}{2}^+$&$1^+$&$D_1^{0\prime}$&$2427(36) $ &$...$&$...$        &$D^{+\prime}_{s1}$&$2459.5(6)$\\  
$\frac{1}{2}^+$&$0^+$&$D_0^{*0}$     &$2300(19)$  &$D_0^{*+}$&$2349(7)$        &$D_{s0}^{*+}$     &$2317.8(5)$\\  
$\frac{1}{2}^-$&$1^-$&$D^{*0}$       &$2006.85(5)$&$D^{*+}$         &$2010.26(5)$ &$D^{*+}_s$        &$2112.2(4)$\\  
$\frac{1}{2}^-$&$0^-$&$D^0$          &$1864.83(5)$&$D^+$            &$1869.65(5)$ &$D^+_s$           &$1968.34(7)$\\  \hline\hline
\end{tabular}
\label{tab:1}
\end{center}
\end{table}

To fit the mass expansion in Eq.~\eqref{rmasses} to the experiment, 
let us first define the following experimental residual masses for the excited charmed mesons,
 \begin{equation}\label{rm11}
\begin{split}
&m_{D^*_0}=340.4(7.0)~ \text{MeV}, ~~~~~~~~m_{D^*_{0s}}=309.25(50) ~\text{MeV},\\
&m_{D^\prime_1}=418(36) ~\text{MeV}, ~~~~~~~~~~~m_{D^\prime_{1s}}= 450.95(60)~\text{MeV},\\
&m_{D_1}=413.4(1.2)~ \text{MeV}, ~~~~~~~~m_{D_{1s}}= 526.555(70)~ \text{MeV},\\
&m_{D^*_2}=454.50(68) ~\text{MeV},~~~~~~~m_{D^*_{2s}}=560.55(80)~\text{MeV},
\end{split}
\end{equation}
which are measured from the nonstrange vector charmed meson mass, $m_{D^*}$. The nonstrange flavor index $n$ in Eq.~\eqref{rm11} is suppressed. 
By fitting our one-loop mass formula [Eq.~\eqref{rmasses}] to the corresponding experimental spectrum [Eq.~\eqref{rm11}], 
one gets   
\begin{align}\label{dds}
&\eta_S=506(19)~ \text{MeV}, ~~~~~\xi_S=65(26)~ \text{MeV},\\\label{lls}
&L_S=29(30)~ \text{MeV},  ~~~~~F_S=49 (39)~ \text{MeV},\\\label{dt}
&\eta_T=637(1)~ \text{MeV},  ~~~~~~\xi_T=140(1)~ \text{MeV},\\\label{lt}
&L_T=184(1)~ \text{MeV},  ~~~~~F_T=45(2)~ \text{MeV},
\end{align}
where the associated uncertainties involve the experimental errors on charmed meson masses and couplings ($g$ and $h$) and the error on $g^\prime$ from LQCD. The errors are dominated by the uncertainty in the $D^*_0$ and $D^\prime_1$ masses. 
 
Using Eqs.~\eqref{mm2},~\eqref{lls}, and \eqref{lt}, one can fix the $SU(3)$-breaking coefficients, 
\begin{equation}
\begin{split}\label{las}
a_S&=0.23(24),~~~~~~~\Delta^{(a)}_S=0.39(31),~~~~~~~a_T=1.4592(87),~~~~~~~\Delta^{(a)}_T=0.355(16).
\end{split}
\end{equation}

To fix the other coefficients ($\delta_A$, $\Delta_A$, $\sigma_A$, $\Delta^{(\sigma)}_A$), the combinations $\eta_A$ and $\xi_A$  must be extracted at different light quark masses as the 
experimental information [see Eqs.~\eqref{dds} and  \eqref{dt}] is not enough to disentangle
chirally symmetric terms from the $SU(3)$ symmetric one; see Eq.~\eqref{cof}. 
For our purpose, the continuum lattice results on the  charmed meson spectroscopy that are computed at different light quark masses 
are required. Such findings are made available in \cite{lattice}. In our fit, we consider the results that are obtained using the lightest pion masses ($m_\pi\lesssim 250$ MeV), which are within the validity of ChPT. Such data are given in ensembles D15.48 ($m_n=5$ MeV, $m_s=382$ MeV, $m_\pi=224$ MeV) and D20.48 ($m_n=6$ MeV, $m_s=382$ MeV, $m_\pi=257$ MeV) of Ref. \cite{lattice}.
In Table \ref{tab:2}, the continuum charmed meson masses at nonphysical pion masses in each ensemble are
presented. To make the continuum extrapolation,  strategy 3 explained in \cite{lattice} is employed.
As the discretization errors are negligible, one can safely use the mass relations
$((2m^2_K-m^2_{\pi})_{\text{phys}}+m^2_{\pi,L})/2$ and 
$(2(2m^2_K-m^2_{\pi})_{\text{phys}}+m^2_{\pi,L})/3$ to obtain $m^2_K$ and $m^2_\eta$, respectively, where
the subscript $L$ means the lattice measured pion mass.
These mass relations are valid as the calculations in \cite{lattice} performed at the physical value of the strange valence quark mass; i.e., the physical value of  $2m^2_K-m^2_{\pi}$ is reproduced using $m_{K,L}$ and $m_{\pi,L}$ measured in each ensemble. In leading order ChPT, the quantity $2m^2_K-m^2_{\pi}$
gives the strange light quark mass and is not sensitive  to the nonstrange light quark mass.
The errors associated with the lattice calculations of the charmed meson masses are negligible at our level of precision.

\begin{table}[h!]
\def\arraystretch{1.5}
\begin{tabular}[t]{|c|ccccccccccc|}
\toprule
\hline
Ensemble&$m_{D^*}$&$m_{D_s}$ & $m_{D^*_s}$&$m_{D^*_0}$&$m_{D^\prime_1}$&$m_{D^*_{s0}}$& $m_{D^\prime_{s1}}$& $m_{D_1}$&$m_{D_{s1}}$&$m_{D_2}$&$m_{D_{s2}}$\\
\hline
\bottomrule
D15.48& 2029.0(7.0)& 1962.6(2.8)& 2119.3(3.8) & 2351(10) & 2490(15) & 2400(11) & 2565(10)& 
2634(22) & 2624(22) & 2747(30) & 2742(25)\\
D20.48 & 2030.0(7.1)&1959.9(2.8)&2117.7(3.9)&2364(10) & 2503(15) & 2404(11) & 2570(10)& 
2636(22) & 2627(23) & 2754(31) & 2745(25) \\
\hline
\end{tabular}
\caption{The strategy 3 illustrated in \cite{lattice} is used to obtain the above continuum masses, which are given in MeV units. In Ref. \cite{lattice}, 
the ground-state mass $m_D$ was used to tune the charm quark mass. In our fit, we use the experimental value given in Table \ref{tab:1} for this nonmeasured lattice mass.}
\label{tab:2}
\end{table}
By fitting the mass formula in Eq.~\eqref{rmasses} to the lattice results [Table \ref{tab:2}] on the residual masses, 
one finds 
\begin{align}\nonumber
\text{D15.48}:\\\label{ln1}
&\eta_S=788(12), ~~~~~\xi_S=132(15), ~~~\eta_T=1022(16),
~~~~\xi_T=228(29),\\\label{ln2}
&L_S=272(20),  ~~~~F_S=-3(27), ~~~L_T=158(30),~~~~~F_T=138 (56),\\\nonumber
\text{D20.48}:\\\label{ln3}
&\eta_S=811(12), ~~~~~\xi_S=130(15), ~~~\eta_T=1037(16),
~~~~\xi_T=237 (29),\\\label{ln4}
&L_S=259(20),  ~~~~F_S=2(27),~~~~~L_T=162(31),~~~~~F_T=136(56),
\end{align}
which are given in MeV units. The associated uncertainties include the experimental errors on the couplings and errors on the charmed meson masses from LQCD.

What matters to us from the lattice fit [Eqs.~\eqref{ln1}--\eqref{ln4}] is that $\eta_A$ and $\xi_A$ are extracted at different (nonphysical) light quark masses. Therefore, using experimental [Eqs.~\eqref{dds} and \eqref{dt}] and lattice
[Eqs.~\eqref{ln1} and \eqref{ln3}] results, one can now separate the chiral symmetric terms from those that respect $SU(3)$ symmetry as shown in  
Eq.~\eqref{cof}.
To do so, a constrained fitting procedure \cite{prior} is utilized. In the fit,
the extracted values in Eq.~\eqref{las} are used as priors on the coefficients $a_A$ and  $\Delta^{(a)}_A$. For the coefficients $\delta_A$, $\Delta_A$, $\sigma_A$, and $ \Delta^{(\sigma)}_A$, broad priors are used. We choose $0\pm 1000$ MeV ($0\pm 1000$) as priors on  $\delta_A$ and $\Delta_A$ ($\sigma_A$ and $ \Delta^{(\sigma)}_A$).
Performing a least chi-squared fit, one gets
\begin{equation}
\begin{split}\label{l113}
\delta_S&=346(30)~\text{MeV},~~~~~~~ \Delta_S=29(40)~\text{MeV},
~~~~~\Delta^{(\sigma)}_S=0.13(15),~~~~~~~~~~~\sigma_S=1.08(11),\\
\delta_T&=426(6)~\text{MeV},~~~~~~~~ \Delta_T=90(11)~\text{MeV},
~~~~~\Delta^{(\sigma)}_T=0.243(81),~~~~~~~~~~\sigma_T=1.047(45),
\end{split}
\end{equation}
where the uncertainties on the above values include the experimental errors of charmed meson masses and coupling constants and errors from 
lattice data on charmed meson masses. 

Clearly, the extracted values given in Eqs.~\eqref{las} and \eqref{l113} for the coefficients that appear in the effective chiral Lagrangian are consistent with the perturbative expansion of the theory. 
By fitting the counterterms, our one-loop mass expressions given in Eq.~\eqref{masses} can be used to study several mass splittings within excited charmed mesons. As an illustration, let us use the theory to compute the 
hyperfine splitting,
\begin{align}\label{sss}
&m_{D^*_{2s}}-m_{D_{1s}}=\Delta_T+\Delta_T^{(a)}m_s+\Delta_T^{(\sigma)}\overline{m}+\Sigma_{D^*_{2s}}-\Sigma_{D_{1s}}.
\end{align}
Using our results, one gets $34(16)$ MeV for this hyperfine splitting, which  agrees well with the observed value,
$m_{D^*_{2s}}-m_{D_{1s}}=33.99(80)$ MeV; see Table \ref{tab:1}.
Well-measured experimental and lattice data on charmed meson masses are necessary to reduce the uncertainties on the coefficients [Eqs.~\eqref{las} and \eqref{l113}] and hence the predicted
hyperfine splitting. Our results can also be used to predict the analog bottom meson states and this is left for future work. 

\section{Appendix}
Here we present the explicit expressions for  the self-energies of the
excited charmed mesons 
\begin{equation}
\begin{split}
 \Sigma_{D^*_0}&=\frac{g^{\prime 2}}{4 f^2}\left[ 3K_1(m_{D^\prime_1}-m_{D^*_0},m_{\pi})+\frac{1}{3}K_1(m_{D^\prime_1}-m_{D^*_0},m_{\eta})+2 K_1(m_{D^\prime_{1s}}-m_{D^*_0},m_K) \right]\\[2ex]
&+\frac{h^2}{4 f^2}\left[ 3K_2(m_D-m_{D^*_0},m_{\pi})+\frac{1}{3}K_2(m_D-m_{D^*_0},m_{\eta})+2 K_2(m_{D_s}-m_{D^*_0},m_K)\right] \\[2ex]
&+\frac{h^{\prime 2}}{4 f^2}\left[ \frac{2}{3}\left(3K_1(m_{D_1}-m_{D^*_0},m_{\pi})+\frac{1}{3}K_1(m_{D_1}-m_{D^*_0},m_{\eta})+2 K_1(m_{D_{1s}}-m_{D^*_0},m_K)\right)\right],
\end{split}
\end{equation}
\begin{equation}
\begin{split}
\Sigma_{D^*_{0s}}&=\frac{g^{\prime 2}}{4 f^2}\left[4 K_1(m_{D^\prime_1}-m_{D^*_{0s}},m_K)+\frac{4}{3}K_1(m_{D^\prime_{1s}}-m_{D^*_{0s}},m_{\eta}) \right]\\[2ex]
&+\frac{h^2}{4 f^2}\left[4 K_2(m_D-m_{D^*_{0s}},m_K)+\frac{4}{3}K_2(m_{D_s}-m_{D^*_{0s}},m_{\eta}) \right]\\
&+\frac{h^{\prime 2}}{4 f^2}\left[\frac{2}{3}\left(4 K_1(m_{D_1}-m_{D^*_{0s}},m_K)+\frac{4}{3}K_1(m_{D_{1s}}-m_{D^*_{0s}},m_{\eta})\right)\right],
\end{split}
\end{equation}
\begin{equation}
\begin{split}
 \Sigma_{D^\prime_1}&=\frac{g^{\prime 2}}{4 f^2}\left[\frac{1}{3}\left(3K_1(m_{D^*_0}-m_{D^\prime_1},m_{\pi})+\frac{1}{3}K_1(m_{D^*_0}-m_{D^\prime_1},m_{\eta})+2K_1(m_{D^*_{0s}}-m_{D^\prime_1},m_K)\right) \right]\\[2ex]
&+\frac{g^{\prime 2}}{4 f^2}\left[\frac{2}{3}\left( 3K_1(0,m_{\pi})+\frac{1}{3}K_1(0,m_{\eta})+2K_1(m_{D^\prime_{1s}}-m_{D^\prime_1},m_K) \right)\right]\\[2ex]
&+\frac{h^2}{4 f^2}\left[3K_2(m_{D^*}-m_{D^\prime_1},m_{\pi})+\frac{1}{3}K_2(m_{D^*}-m_{D^\prime_1},m_{\eta})+2 K_2(m_{D^*_s}-m_{D^\prime_1},m_K) \right]\\
&+\frac{h^{\prime 2}}{4 f^2}\left[\frac{1}{9}\left(3K_1(m_{D_1}-m_{D^\prime_1},m_{\pi})+\frac{1}{3}K_1(m_{D_1}-m_{D^\prime_1},m_{\eta})+2K_1(m_{D_{1s}}-m_{D^\prime_1},m_K)\right) \right]\\[2ex]
&+\frac{h^{\prime 2}}{4 f^2}\left[\frac{5}{9}\left(3K_1(m_{D^*_2}-m_{D^\prime_1},m_{\pi})+\frac{1}{3}K_1(m_{D^*_2}-m_{D^\prime_1},m_{\eta})+2K_1(m_{D^*_{2s}}-m_{D^\prime_1},m_K)\right) \right],
\end{split}
\end{equation}
\begin{equation}
\begin{split}
\Sigma_{D^\prime_{1s}}&=\frac{g^{\prime 2}}{4 f^2}\left[\frac{1}{3}\left(4 K_1(m_{D^*_0}-m_{D^\prime_{1s}},m_K)+\frac{4}{3}K_1(m_{D^*_{0s}}-m_{D^\prime_{1s}},m_{\eta}) \right)\right]\\[2ex]
&+\frac{g^{\prime 2}}{4 f^2}\left[\frac{2}{3}\left(4 K_1(m_{D^\prime_1}-m_{D^\prime_{1s}},m_K)+\frac{4}{3}K_1(0,m_{\eta}) \right)\right]\\[2ex]
&+\frac{h^2}{4 f^2}\left[4 K_2(m_{D^*}-m_{D^\prime_{1s}},m_K) +\frac{4}{3}K_2(m_{D^*_s}-m_{D^\prime_{1s}},m_{\eta})\right]\\
&+\frac{h^{\prime 2}}{4 f^2}\left[\frac{1}{9}\left(4 K_1(m_{D_1}-m_{D^\prime_{1s}},m_K)+\frac{4}{3}K_1(m_{D_{1s}}-m_{D^\prime_{1s}},m_{\eta}) \right)\right]\\[2ex]
&+\frac{h^{\prime 2}}{4 f^2}\left[\frac{5}{9}\left(4 K_1(m_{D^*_2}-m_{D^\prime_{1s}},m_K)+\frac{4}{3}K_1(m_{D^*_{2s}}-m_{D^\prime_{1s}},m_{\eta}) \right)\right],
\end{split}
\end{equation}
\begin{equation}
\begin{split}
 \Sigma_{D_1}&=\frac{g^{\prime\prime 2}}{4 f^2}\left[\frac{5}{54} \left(3K_1(m_{D^*_2}-m_{D_1},m_{\pi})+\frac{1}{3}K_1(m_{D^*_2}-m_{D_1},m_{\eta})+2K_1(m_{D^*_{2s}}-m_{D_1},m_K)\right) \right]\\[2ex]
&+\frac{g^{\prime\prime 2}}{4 f^2}\left[\left(\frac{2}{3}\right)\left(\frac{5}{6}\right)^2\left( 3K_1(0,m_{\pi})+\frac{1}{3}K_1(0,m_{\eta})+2K_1(m_{D_{1s}}-m_{D_1},m_K) \right)\right]\\[2ex]
&+\frac{h^{\prime 2}}{4 f^2}\left[\left(\frac{2}{9}\right)\left(3K_1(m_{D^*_0}-m_{D_1},m_{\pi})+\frac{1}{3}K_1(m_{D^*_0}-m_{D_1},m_{\eta})+2 K_1(m_{D^*_{0s}}-m_{D_1},m_K)\right) \right]\\
&+\frac{h^{\prime 2}}{4 f^2}\left[\left(\frac{1}{9}\right)\left(3K_1(m_{D^\prime_1}-m_{D_1},m_{\pi})+\frac{1}{3}K_1(m_{D^\prime_1}-m_{D_1},m_{\eta})+2 K_1(m_{D^\prime_{1s}}-m_{D_1},m_K)\right) \right],
\end{split}
\end{equation}

\begin{equation}
\begin{split}
\Sigma_{D_{1s}}&=\frac{g^{\prime \prime 2}}{4 f^2}\left[\left(\frac{5}{54}\right)\left(4 K_1(m_{D^*_2}-m_{D_{1s}},m_K)+\frac{4}{3}K_1(m_{D^*_{2s}}-m_{D_{1s}},m_{\eta}) \right)\right]\\[2ex]
&+\frac{g^{\prime \prime 2}}{4 f^2}\left[\left(\frac{2}{3}\right)\left(\frac{5}{6}\right)^2\left(4 K_1(m_{D_1}-m_{D_{1s}},m_K)+\frac{4}{3}K_1(0,m_{\eta}) \right)\right]\\[2ex]
&+\frac{h^{\prime 2}}{4 f^2}\left[\left(\frac{2}{9}\right)\left(4 K_1(m_{D^*_0}-m_{D_{1s}},m_K) +\frac{4}{3}K_1(m_{D^*_{0s}}-m_{D_{1s}},m_{\eta})\right)\right]\\[2ex]
&+\frac{h^{\prime 2}}{4 f^2}\left[\left(\frac{1}{9}\right)\left(4 K_1(m_{D^\prime_1}-m_{D_{1s}},m_K) +\frac{4}{3}K_1(m_{D^\prime_{1s}}-m_{D_{1s}},m_{\eta})\right)\right],
\end{split}
\end{equation}

\begin{equation}
\begin{split}
 \Sigma_{D^*_2}&=\frac{g^{\prime\prime 2}}{4 f^2}\left[\left(\frac{1}{18}\right) \left(3K_1(m_{D_1}-m_{D^*_2},m_{\pi})+\frac{1}{3}K_1(m_{D_1}-m_{D^*_2},m_{\eta})+2K_1(m_{D_{1s}}-m_{D^*_2},m_K)\right) \right]\\[2ex]
&+\frac{g^{\prime\prime 2}}{4 f^2}\left[\left(\frac{4}{3}\right)\left( 3K_1(0,m_{\pi})+\frac{1}{3}K_1(0,m_{\eta})+2K_1(m_{D^*_{2s}}-m_{D^*_2},m_K) \right)\right]\\[2ex]
&+\frac{h^{\prime 2}}{4 f^2}\left[\left(\frac{1}{3}\right)\left(3K_1(m_{D^\prime_1}-m_{D^*_2},m_{\pi})+\frac{1}{3}K_1(m_{D^\prime_1}-m_{D^*_2},m_{\eta})+2 K_1(m_{D^\prime_{1s}}-m_{D^*_2},m_K)\right) \right],
\end{split}
\end{equation}

\begin{equation}
\begin{split}
\Sigma_{D^*_{2s}}&=\frac{g^{\prime \prime 2}}{4 f^2}\left[\left(\frac{1}{18}\right)\left(4 K_1(m_{D_1}-m_{D^*_{2s}},m_K)+\frac{4}{3}K_1(m_{D_{1s}}-m_{D^*_{2s}},m_{\eta}) \right)\right]\\[2ex]
&+\frac{g^{\prime\prime 2}}{4 f^2}\left[\left(\frac{4}{3}\right)\left(4 K_1(m_{D^*_2}-m_{D^*_{2s}},m_K)+\frac{4}{3}K_1(0,m_{\eta}) \right)\right]\\[2ex]
&+\frac{h^{\prime 2}}{4 f^2}\left[\left(\frac{1}{3}\right)\left(4 K_1(m_{D^\prime_1}-m_{D^*_{2s}},m_K) +\frac{4}{3}K_1(m_{D^\prime_{1s}}-m_{D^*_{2s}},m_{\eta})\right)\right].
\end{split}
\end{equation}
The chiral loop integrals $K_1(\omega,m)$ and $K_2(\omega,m)$ are
\cite{Alhakami}
\begin{equation}
\begin{split}
K_1(\omega, m)&=\frac{1}{16 \pi^2}\left[(-2\omega^3+3m^2\omega)\mathrm{ln}\left(\frac{m^2}{\mu^2}\right)-4(\omega^2- m^2)F(\omega,m)+\frac{16}{3}\omega^3-7\omega\, m^2\right],\\[2ex]
K_2(\omega, m)&=\frac{1}{16 \pi^2} \left[ (-2\omega^3+ m^2\omega)\mathrm{ln}\left(\frac{m^2}{\mu^2}\right)-4\omega^2 F(\omega,m)+4\omega^3- \omega\, m^2\right],
\end{split}
\end{equation}
 renormalized in the $\mathrm{\overline{MS}}$ scheme
 and the function $F(\omega,m)$ is defined as 
\begin{equation}\label{F-function}
 F(\omega,m)=\left\{ \begin{array}{c c}-\sqrt{m^2-\omega^2} \cos^{-1}(\frac{\omega}{m}), & \mbox{$m^2>\omega^2,$} \\[2ex]
                                 \sqrt{\omega^2-m^2}[i \pi-\cosh^{-1}(-\frac{\omega}{m})], & \mbox{$\omega<-m,$}\\[2ex]
 \sqrt{\omega^2-m^2}\cosh^{-1}(\frac{\omega}{m}), & \mbox{$\omega>m.$}
                                \end{array} \right.
\end{equation}

\end{document}